\documentclass[letter]{aa} 
\usepackage{natbib}
\usepackage{graphicx}
\usepackage{url}

\newcommand{\kms}{$\mathrm {km s}^{-1}$}

\newcommand{\teff}{T$_{\rm eff}$}

\usepackage{txfonts}
\begin{document}
   \title{The lithium content of $\omega$~Centauri}   

   \subtitle{New clues to the cosmological Li problem from old stars in external
galaxies\thanks{Based on observations taken at ESO VLT Kueyen telescope (Cerro
Paranal, Chile, program: 079.D-0021A)}}

   \author{L. \,Monaco\inst{1,2}, 
   P. \,Bonifacio\inst{3,4}, 
   L. \,Sbordone \inst{5},\\
   S. \,Villanova\inst{1}, 
         \and
   E. \,Pancino \inst{6}
          }

\institute{
Universidad de Concepci\'on,
Casilla 160-C, Concepci\'on, Chile
\and
European Southern Observatory, Casilla 19001, Santiago, Chile
\and
GEPI, Observatoire de Paris, CNRS, Universit´e Paris Diderot ; Place Jules
Janssen, 92190 Meudon, France
\and
Istituto Nazionale di Astrofisica --
Osservatorio Astronomico di Trieste, Italy
via G. B. Tiepolo 11 34143 Trieste, Italy
\and
Max Planck Institut for Astrophysics
Karl-Schwarzschild-Str. 1
85741 Garching, Germany 
\and
Istituto Nazionale di Astrofisica --
Osservatorio Astronomico di Bologna, 
Via Ranzani 1, 40127, Bologna, Italy
}

\authorrunning{Monaco et al.}
\mail{lmonaco@eso.org}

\titlerunning{The lithium content of $\omega$~Centauri}

   \date{Received; accepted}

 
  \abstract
  %
   {A discrepancy has emerged between the cosmic lithium abundance inferred by
   the WMAP satellite measurement coupled with the prediction of the standard
   big-bang nucleosynthesis theory, and the constant Li abundance measured in
   metal-poor halo dwarf stars (the so-called {\em Spite plateau}). Several
   models are being proposed to explain this discrepancy, involving either new
   physics, in situ depletion, or the efficient depletion of Li in the pristine Galaxy
   by a generation of massive first stars. The realm of possibilities may be
   narrowed considerably by observing stellar populations in different galaxies,
   which  have experienced different evolutionary histories.}
  %
  %
   {The $\omega$~Centauri stellar system is commonly considered as the remnant
   of a dwarf galaxy accreted by the Milky Way. We investigate the lithium
   content of a conspicuous sample of unevolved stars in this object.}
  %
  %
   {We obtained moderate resolution (R=17\,000) spectra for 91
   main-sequence/early sub-giant branch (MS/SGB) $\omega$~Cen stars using the
   FLAMES-GIRAFFE/VLT spectrograph. Lithium abundances were derived by matching
   the equivalent width of the Li~{\sc i} resonance doublet at 6708~\AA\, to the
   prediction of synthetic spectra computed with different Li abundances.
   Synthetic spectra were computed using the SYNTHE code along with ATLAS--9
   model atmospheres. The stars effective temperatures are derived by fitting
   the wings of the H$_\alpha$ line with synthetic profiles.}
  %
  %
   {We obtain a mean content of A(Li)=2.19$\pm$0.14~dex for $\omega$ Centauri
   MS/SGB stars. This is comparable to what is observed in Galactic halo field
   stars of similar metallicities and temperatures.}
  %
  %
   {The Spite plateau seems to be an ubiquitous feature of old, warm metal-poor
   stars. It exists also in external galaxies, if we accept the current view
   about the origin of $\omega$~Cen. This implies that the mechanism(s) that
   causes the ``cosmological  lithium problem'' may be the same in the Milky Way
   and other galaxies.}



   \maketitle
%

\section{Introduction}

According to standard cosmology, light elements ($^2$H, $^3$He, $^4$He and
$^7$Li) were synthesized, starting from protons, during the initial hot and
dense phase of the evolution of the Universe. The yields of this primordial
nucleosynthesis depend on the baryon-to-photon ratio $\eta$, a cosmological
parameter not constrained by first principles. 

The remarkable constancy of the Li abundance, observed in warm, unevolved
metal-poor stars of different effective temperature and metallicity --- the
so-called ``Spite plateau" \citep[][]{spitea,spiteb} has long been interpreted
as a signature of big bang nucleosynthesis and a tool for measuring the
baryonic density of the Universe \citep[see][]{spite10,gary}.

However, the baryonic density $\eta$ has been measured with unprecedented
precision, from the fluctuations of the cosmic microwave background, by the WMAP
satellite \citep[][hereafter C08]{cyburt08}. Coupled with the prediction of the
standard  big bang nucleosynthesis (SBBN) theory, this measurement implies a
primordial Li abundance  a factor of three to four higher than the Spite
plateau. This discrepancy is often referred to as the ``cosmological lithium
problem''.

Many solutions for this discrepancy have been  proposed, including new physics
at the time of the big bang \citep[][]{jed04,jed06,jittoh,hisano09}, astration
in the pristine Galaxy \citep[][hereafter P06]{piau06}, and turbulent diffusion
to deplete lithium in the stellar atmospheres \citep[][]{ric05}. 

A fresh look at the problem can be afforded by the study of lithium in
metal-poor populations of external galaxies. Theories such as that of P06 can
be  immediately tested and also theories that invoke stellar phenomena can be
constrained by the observation of systems that have experienced different star
formation histories. 

The $\omega$~Centauri stellar system is commonly considered as the remnant of a
dwarf galaxy accreted by the Milky Way
\citep[e.g.,][]{zinnecker88,freeman93,carraro00}. The complexity of its
color-magnitude diagram \citep[e.g.][]{lee99,pancino00,antonio05,sandro07}
clearly testify the existence of multiple stellar populations, spanning a
sizeable range in metallicities ($-0.6<$[Fe/H]$<-2.1$,
\citealt[][]{pancino02,antonio05b,sandro07}). 

In this Letter, we present the first measurements of the lithium content in a
conspicuous sample of main-sequence and early sub-giant branch (MS/SGB) stars in
$\omega$~Cen.

%

\section{Observations and data reduction}

\begin{figure}
  \centering
  \includegraphics[width=9cm]{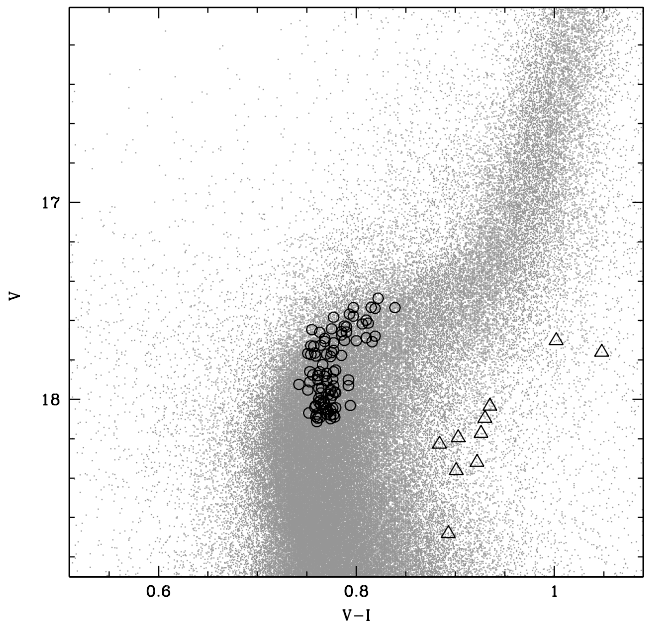} 

     \caption{Target stars are plotted on top of the  $\omega$~Centauri V {\it
     versus} V -- I color-magnitude diagram. Open circles mark MS/SGB stars,
     open triangles SGB-a stars.} 

	\label{cmd}
  \end{figure}

We selected targets at the turn-off and sub-giant branch level of $\omega$~Cen,
mainly using the high precision FORS/VLT photometry of \citet[][hereafter
S05]{antonio05} and data from the spectroscopic survey of  \citet[][hereafter
V07]{sandro07}.  Ten targets were selected to trace the so-called SGB-a (S05).
Target stars are shown in Fig.\,\ref{cmd}, superimposed on the wide field
photometry of \citet[][hereafter B09]{bellini09}. The MS targets selected from
S05 cover a more central area. For these stars, we use in Fig.~\ref{cmd} the
original S05 photometry corrected for the zeropoint difference between B09 and
S05 at the magnitude level of interest.

Observations were taken on three nights of 27-29 April 2007 using the FLAMES/VLT
spectrograph at ESO Paranal. The fibers in Medusa mode fed the GIRAFFE
spectrograph, configured in the HR15n setting, which covers both H$_\alpha$ and
the Li\,{\sc i} resonance doublet at 670.8\,nm at a resolution of 17\,000. The
same plate configuration was  observed for all three nights, with integration
times between one hour and slightly over two hours.  One additional plate
configuration was observed, which has a partial overlap with the main plate
configuration.  We thus achieved a total integration time of between  17 and 19
hours for most of our MS/SGB and SGB-a stars.

%

Frames were processed using version 2.13 of the FLAMES/GIRAFFE data reduction
pipeline\footnote{\url{http://girbldrs.sourceforge.net/}}. A total of 17 fibers
were allocated for sky subtraction on each plate. The average of the sky fibers
closer to each star was subtracted from the science spectra. The standard
IRAF\footnote{IRAF is distributed by the National Optical Astronomy
Observatories, which is operated by the association of Universities for Research
in Astronomy, Inc., under contract with the National Science Foundation.} task
{\it fxcor} was employed to measure the star's radial velocities by
cross-correlating the spectra with a synthetic one of similar atmospheric
parameters. Correction to the heliocentric system were computed using the IRAF
task {\it rvcorrect} and applied to the observed radial velocities. After being
reduced to rest frames, multiple spectra of the same target were finally
averaged. Typical errors of $\sim$1.2~kms$^{-1}$ are derived from repeated
measurements of the stellar radial velocity. We end-up with a total of 178
cluster members providing a mean cluster heliocentric velocity and dispersion of
v$_\odot$=233.8~kms$^{-1}$ and $\sigma$=10.9~kms$^{-1}$. These values are in
good agreement with FLAMES-GIRAFFE/VLT measures presented by \citet[][hereafter
P07, v$_\odot$=233.4~kms$^{-1}$; $\sigma$=13.2~kms$^{-1}$]{pancino07} for a
sample of 649 $\omega$~Cen giants.  Given the significant cluster rotation, a
more detailed comparison should take into account the number of stars sampled in
each cluster region. 

\begin{figure}
  \centering
  \includegraphics[width=9cm]{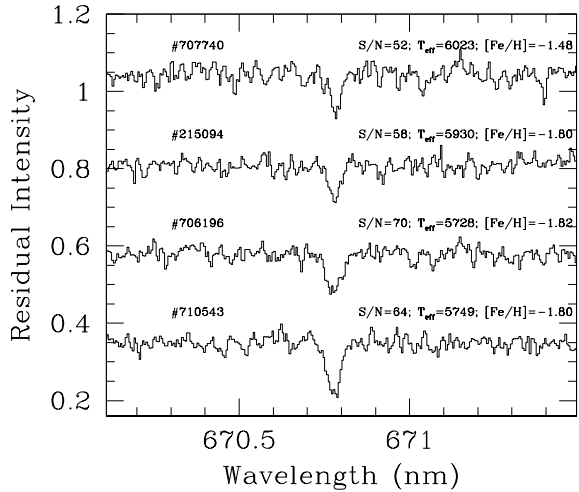} 

     \caption{Summed spectra for a subsample of stars with measured lithium
     abundances. The stellar \teff\, and [Fe/H] in addition to the spectrum
     signal-to-noise ratio (S/N) are indicated.} 

	\label{spec}
  \end{figure}

In the following, we consider only stars with repeated observations. All of them
(91 MS/SGB, 10 SGB-a) have radial velocities within 3~$\sigma$ of the cluster
motion as derived above. After data reduction, we obtained averaged spectra for
these stars with signal-to-noise ratios (S/N) in the range 30 to 90 with a mean
around 60. Stars without repeated observations have a too low S/N, which make
them unhelpful for the present analysis. Spectra for a subsample of stars with
measured lithium abundance are presented in Fig.~\ref{spec}.

%

\section{Abundance analysis}

We derived stellar lithium abundances from the equivalent width (EW) of the 
Li~{\sc i} resonance doublet at $\sim$6708~\AA\,. The EWs were measured by
fitting synthetic profiles, as done in \citet[][]{bonifacio02}. When we could
not detect the Li line, we estimated an upper limit given by 2$\sigma_{EW}$,
where $\sigma_{EW}$ was estimated using the Cayrel formula \citep[][]{cayrel88}.
Abundances were then derived by iteratively computing synthetic profiles until
the EW of the Li doublet matched the measurements. For each star, synthetic
spectra are calculated using the SYNTHE code along with an appropriate
one-dimensional ATLAS--9 model atmosphere \citep[][]{sbordone04,Sb05,kurucz05}.
The effective temperature of the stars was determined by fitting the wings of
the H$_\alpha$ line. The theoretical  profiles were computed with a modified
version of the BALMER code\footnote{The original version provided by R.\,L.
Kurucz is available  at \url{http://kurucz.harvard.edu/}}, which uses the
self-broadening theory of \citet[][]{barklem00,barklem00b} and the Stark
broadening given by \citet[][]{stehle99}. We assumed log~g~=~4.0 and a
metallicity of [Fe/H]=-1.5 for all stars, thus ignoring the dependence of the
Balmer line profiles on both metallicity and surface  gravity. A microturbulent
velocity of 1 \kms\ was assumed. This, like the assumed surface gravity and
metallicity, have effects of a few hundredths of dex on the derived Li
abundance. This is totally negligible in the current context. Our S/N ratios are
sufficiently high to ensure that the error in the Li abundances is totally
dominated by the uncertainty in the effective temperatures. The latter is on the
order of 150\,K and is dominated by the uncertainty  in the correction of the
blaze function of GIRAFFE. Our estimated internal uncertainty in the Li
abundance is 0.1\,dex.

For none of the SGB-a stars did we detect the Li doublet. For these stars, we
derived upper limits in the range
A(Li)\footnote{A(Li)=log$\frac{n(Li)}{n(H)}$+12.00.}$<$0.9-1.5~dex. Owing to
their cool temperature (\teff$\leq$5650\,K), some level of lithium depletion is,
however, actually expected for these stars. From a sample of 91 RGB/SGB stars,
we could measure the Li abundance of 52, and for the remaining 39 we provide
upper limits. The mean value was found to be A(Li)=2.19 with a dispersion of
0.14\,dex. When we, instead, adopt effective temperatures based on the V--I
color \citep[][]{alonsodwarf96} the mean Li abundance and dispersion we obtain
are identical to the above figures within the errors, i.e. A(Li)=2.17$\pm$0.12.
We used a reddening of E(B-V)=0.11 \citep[][]{lub02}.

The spectral region covered by our observations is not rich in metallic lines.
Nevertheless we used the available Fe~{\sc i} and Ca~{\sc i} lines to estimate
the  metallicity of the stars, assuming [Ca/Fe]=+0.4. For the stars in common,
we find a rather good agreement with the results of V07, although we obtain
slightly lower metallicities on the order of $\sim$0.1\,dex. In the online
material we report, coordinates, effective temperatures, [Fe/H], A(Li), and the 
equivalent width of the Li line for the target stars.

  \begin{figure}
  \centering
  \includegraphics[width=9cm]{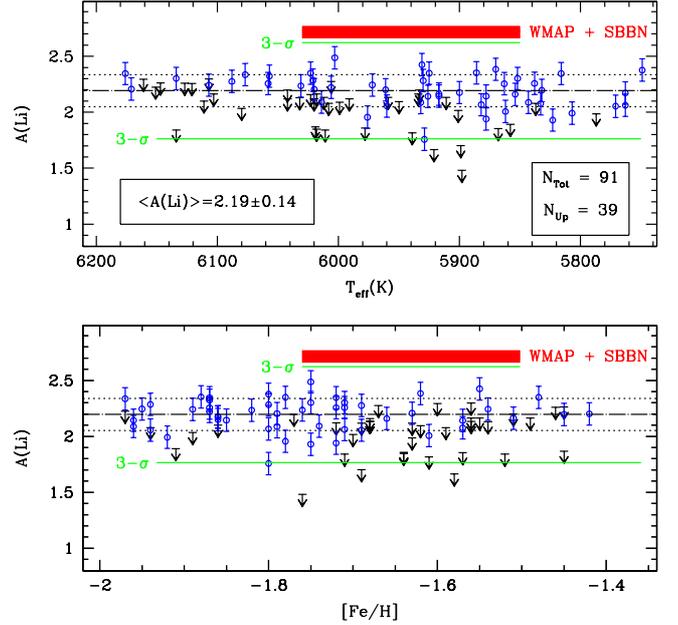}

	\caption{Measured lithium abundances for targets on the main
	sequence/sub-giant branch as a function  of the effective temperature
	(upper panel) and the star  metallicity (lower panel).  The mean Li
	abundance and dispersion derived for $\omega$~Cen, the number of stars
	analyzed (N$_{Tot}$), and the number of stars for which we derived upper
	limits (N$_{Up}$) are indicated in the upper panel. The mean Li
	abundance (dot-dashed line), the 1--$\sigma$ (dotted lines), and 
	3--$\sigma$ (continuous lines) levels from the mean and the primordial
	lithium level implied by  WMAP measures plus SBBN theory (C8, shaded
	area) are also marked for reference.}\label{figteff}

  \end{figure}

%

\section{Discussion and conclusions}

We have either measured or derived upper limits to the Li abundances of 91
MS/SGB stars in the $\omega$~Cen stellar system. We have also estimated upper
limits for 10 stars belonging to the so-called SGB-a. 

Our results for MS/SGB stars are summarized in Fig.~\ref{figteff}. The
$\omega$~Cen stars lie on the Spite plateau, well below the value implied by the
WMAP measurement coupled with the SBBN theory.  Comparison with the Galactic
field stars on the same effective temperature scale \,\citep{S10} shows that
$\omega$~Cen stars occupy the same zone as the Galactic ones, in both the
A(Li)--\teff\ and the A(Li)--metallicity planes. 

There are no obvious trends in Li abundance with temperature (upper panel) or
metallicity (lower panel) and upper limits are not confined to any particular
metallicity.  The average lithium abundance of stars fainter than V=17.8 (i.e.,
MS stars) is, however, slightly lower than that of brighter stars (i.e., early
SGB stars) which are 0.06\,dex richer: A(Li)$_{V<17.8}$=2.22$\pm$0.12 (30
stars); A(Li)$_{V>17.8}$=2.16$\pm$0.17 (22 stars).  This is similar to what is
observed in the globular cluster NGC~6397 and in the field
\citep[][]{ngc6397,charbonnel05} and may indicate a dependence of the Li content
on the evolutionary state. The absence of stars with high Li abundances suggests
that stars in our sample were not (significantly) polluted by lithium produced
in asymptotic giant branch (AGB) stars  by the \citet[][]{cf71} mechanism. 

We have a large number of upper limits, but only 3 stars are at 3$\sigma$ from
the mean lithium abundance and 10 at $2 \sigma$.  If $\omega$~Cen has a He-rich
population\,\citep{norris04,p05}, we can expect this population to be
essentially Li-free, since  at the high temperatures necessary for He
production, Li is  completely destroyed. Higher S/N observations are required to
robustly assess the number of  Li-depleted stars, yet it is clear that if these
stars belong, indeed, to a He-rich population, then a Li-normal, and by
inference also He-normal, population of the same metallicity, also exists. The
Spite plateau of $\omega$~Cen appears indeed to be uniformly populated at all
metallicities sampled in the present analysis. 

Globular clusters are known to display light elements abundance variations  down
to the turn-off \citep[][]{carretta09}. In NGC\,6752, 47\,Tuc and perhaps
NGC\,6397, the Na-O anticorrelation is accompanied in MS by a Na-Li
anticorrelation \citep[][]{pasquini05,bonifacio07,lind09}. Some of the stars for
which the Li line was not detected may indeed be Na-rich. On the other hand, the
spread observed in the measured Li abundances is compatible with the analysis
uncertainties.

Based on the hypothesis that $\omega$~Cen is the remnant of an accreted dwarf
galaxy, we can conclude that:

\begin{itemize}
\item  the Spite plateau also exists in other galaxies;
\item  the mechanism(s) causing the ``cosmological  lithium problem'' is(are)
       the same in the Milky Way and other galaxies.
\end{itemize}

The second conclusion is the simplest and most reasonable scenario we may
envisage at the present stage but it clearly requires confirmation by Li
measurements in additional external galaxies. 

Solutions to the ``cosmological lithium problem'' that propose a special
evolution for the Milky Way halo are disfavored by our results.  To reconcile
the primordial value with the current observed Spite plateau level, the
astration  model (P06) requires one third to one half of the Galactic halo
($\sim 10^9$M$_\odot$) to have been processed through a generation of massive
stars that effectively depleted lithium. However, $\omega$~Cen has a different
type, mass, and has likely experienced a different evolution from the Milky Way.
A fine tuning of the mass fraction processed through massive stars would be
required for this different galaxy to end up with Li abundances similar to those
of Milky Way stars.

In contrast, models based on cosmological simulations predict the formation of
the Milky Way halo by the assembly of a large number of sub-units \citep[see,
e.g.,][]{bullock05}. The similar Li content of old metal-poor stars in the halo
and in $\omega$\,Cen may, indeed, favor the hierarchical merging paradigm. What
fraction of the stellar population in the Galaxy halo could have been built-up
by this mechanism is still, however, a matter of debate. The chemical
composition of the oldest, most metal-poor stars in the Milky Way dwarf
satellites indeed provides conflicting evidence in this respect
\citep[][]{monaco07,aoki09,frebel10}. 

It has been suggested that $\omega$\,Cen may be the remnant of the nucleus of a,
now destroyed, dwarf galaxy that had hosted a globular cluster (GC) at its
center \citep[][]{bellazzini08,carretta10}. $\omega$\,Cen would then be a system
similar to the GC M~54 plus the nucleus (SgrN) of the Sagittarius dwarf
spheroidal galaxy (Sgr dSph) at whose center it lies.  In this scenario, part of
the $\omega$\,Cen stellar population would belong to the former galaxy, and the
remaining part to the GC residing at the nucleus center. While we cannot clearly
identify with either origin, both type of stars would, in any case, be of
extra-galactic origin. As for M\,54 \citep[][]{monaco05,bellazzini08}, the
position of $\omega$\,Cen at the bottom of a galaxy potential well, strongly
implies that the GC originates within the galaxy. 

Our observations also have an impact on theories invoking stellar atmospheric
phenomena, such as diffusion\,\citep{ric05}. It has been shown that diffusion
alone is unable to simultaneously reproduce the level of depletion observed in
field stars for Li, Be and B \citep[][]{boesgaard98}. The inclusion of internal
gravity wave, on the other hand, has proven successful in reproducing both the
solar rotation profile and the lithium abundance {\it vs} age trend observed in
open clusters \citep[][]{ct05}. Unfortunately, similar models are still lacking
for populations\,II stars. Diffusion associated with turbulent mixing has been
proposed as a likely solution to the cosmological lithium problem \citep{korn}.
Recent observations, however, cast serious doubts on this hypothesis
\citep{ngc6397}.

These phenomena are time dependent. Since the age spread of the Galactic halo is
very small, however, models cannot be constrained by the observations of halo
dwarfs at the Spite plateau. The true age spread in $\omega$~Cen and within each
of its subpopulations remains a matter of lively debate. Several
studies\,\citep{hughes04,antonio05b,stanford06} concluded that $\omega$~Cen
enriched itself on a timescale $<$2-5~Gyr, with each subpopulation being
essentially coeval within the uncertainties implied in the analysis
\citep[$\sim$1.5~Gyr, see, e.g.,][]{antonio05b}.  On the other hand,
\citet{johnson09} presented evidence of different degrees of s-process
enrichment among stars in the metal-poor population, which is the one sampled in
the present analysis. This implies a time span of  0.1--3~Gyrs
\,\citep{schaller92} for 1.5-3.0\,M$_\odot$ AGB stars to pollute part of the
metal-poor group, depending on the true mean mass of the AGB population. Further
evidence comes from the relative ages derived from color-magnitude diagrams,
combined with spectroscopic metallicities (V07), which infer an age range of
$\sim$6~Gyr over the whole $\omega$~Cen metallicity range. For the 35 stars in
common with the present study, the age spread is  5.6~Gyr, of which 31 have an
age spread of $\sim$2.5~Gyr.

Coupled with our results, we can conclude that theories invoking time-dependent
phenomena should prescribe a constancy of the lithium abundance (at the level of
the observed dispersion, i.e. 0.14~dex) over a timescale comparable to the
above ranges.

\Online

\begin{table*}
\caption{Basic data for MS/SGB stars studied in this paper. For each star, we report ID, coordinates, the Li line EWs and errors, and the lithium abundances. The spectra signal-to-noise, H$_\alpha$-based effective temperatures, and the estimated [Fe/H] and its errors are also reported. For three stars, no [Fe/H] was estimated. These stars were not plotted in the bottom panel of Fig.\ref{figteff}.}\label{abuntable}
\begin{center}								     
\begin{tabular}{c|cc|ccr|c|c|cc}  					     
\hline									     
ID & RA  & Dec & EW & $\epsilon$ (EW) & A(Li) & S/N & \teff (K) & [Fe/H] & $\epsilon$ ([Fe/H])  \\ 	
\hline 								             
200876 &13:26:36.82 &-47:35:55.49 &43.3 & 2.7  &    2.17 &84 &5763 &-1.86 & 0.20\\
202055 &13:26:50.23 &-47:35:35.57 &32.3 & 3.1  &    2.15 &72 &5917 &-1.85 & 0.04\\
212980 &13:26:45.74 &-47:32:58.26 &---- & ---  & $<$1.86 &67 &5978 &-1.64 & 0.17\\
213239 &13:26:37.69 &-47:32:55.16 &54.2 & 6.8  &    2.42 &33 &5931 &-1.55 & 0.40\\
215094 &13:26:25.00 &-47:32:29.91 &41.4 & 3.9  &    2.28 &58 &5930 &-1.80 & 0.24\\
215324 &13:26:31.64 &-47:32:26.92 &---- & ---  & $<$1.85 &59 &6018 &-1.64 & 0.22\\
215862 &13:26:44.65 &-47:32:19.14 &---- & ---  & $<$1.48 &69 &5898 &-1.76 & 0.11\\
216091 &13:26:39.09 &-47:32:16.33 &---- & ---  & $<$2.04 &57 &6080 &-1.89 & 0.40\\
216284 &13:26:58.90 &-47:32:13.07 &---- & ---  & $<$2.13 &55 &5959 &-1.56 & 0.13\\
216434 &13:26:46.08 &-47:32:11.40 &---- & ---  & $<$1.84 &52 &6134 &-1.52 & 0.12\\
216941 &13:26:40.28 &-47:32:04.95 &---- & ---  & $<$2.29 &67 &6161 &-1.60 & 0.22\\
217251 &13:26:30.08 &-47:32:00.75 &28.0 & 3.8  &    2.09 &59 &5932 &-1.79 & 0.18\\
217350 &13:26:56.03 &-47:31:58.72 &---- & ---  & $<$1.87 &54 &6019 &-1.45 & 0.18\\
217836 &13:26:54.36 &-47:31:51.95 &---- & ---  & $<$1.85 &54 &5868 &-1.57 & 0.15\\
218599 &13:26:28.06 &-47:31:42.16 &---- & ---  & $<$2.20 &63 &5933 &-1.77 & 0.08\\
218941 &13:26:39.40 &-47:31:37.41 &---- & ---  & $<$2.26 &54 &6147 &-1.45 & 0.22\\
219646 &13:26:23.86 &-47:31:28.38 &37.6 & 5.0  &    2.34 &45 &6077 &-1.97 & 0.39\\
219921 &13:26:48.18 &-47:31:24.39 &13.9 & 3.5  &    1.76 &64 &5929 &-1.80 & 0.38\\
220173 &13:26:31.33 &-47:31:21.29 &32.2 & 4.5  &    2.30 &49 &6134 &-1.71 & 0.27\\
220472 &13:26:39.15 &-47:31:17.16 &26.4 & 4.2  &    2.01 &53 &5862 &-1.61 & 0.13\\
220539 &13:26:51.79 &-47:31:16.01 &31.9 & 3.6  &    2.08 &63 &5833 &-1.57 & 0.11\\
220543 &13:26:25.51 &-47:31:16.60 &---- & ---  & $<$1.99 &54 &5787 &-1.63 & 0.11\\
221047 &13:27:01.15 &-47:31:09.23 &---- & ---  & $<$2.23 &64 &6151 &-1.97 & 0.19\\
221944 &13:26:27.77 &-47:30:57.94 &31.1 & 5.0  &    2.20 &45 &6020 &-1.79 & 0.05\\
224278 &13:26:46.11 &-47:30:25.83 &---- & ---  & $<$2.16 &68 &6018 &-1.49 & 0.02\\
225188 &13:26:30.13 &-47:30:13.63 &24.9 & 3.1  &    2.09 &71 &6014 &-1.74 & 0.19\\
226046 &13:27:05.23 &-47:30:01.08 &---- & ---  & $<$2.09 &65 &5999 & ---- & ----\\
228016 &13:26:25.53 &-47:29:34.41 &27.9 & 5.0  &    1.99 &45 &5807 &-1.92 & 0.21\\
228199 &13:26:26.89 &-47:29:31.87 &29.0 & 3.6  &    2.07 &62 &5882 &-1.80 & 0.32\\
232117 &13:26:24.07 &-47:28:37.49 &---- & ---  & $<$2.02 &52 &5901 &-1.70 & 0.17\\
232161 &13:27:01.29 &-47:28:35.98 &---- & ---  & $<$1.70 &87 &5899 &-1.69 & 0.02\\
233415 &13:27:08.78 &-47:28:17.85 &---- & ---  & $<$2.10 &72 &5950 &-1.62 & 0.34\\
234553 &13:26:24.61 &-47:28:03.58 &---- & ---  & $<$2.12 &56 &6020 &-1.63 & 0.18\\
234927 &13:26:26.85 &-47:27:58.39 &---- & ---  & $<$2.08 &68 &5837 &-1.94 & 0.28\\
238398 &13:27:03.93 &-47:27:09.94 &---- & ---  & $<$2.28 &58 &6006 &-1.67 & 0.16\\
238851 &13:26:58.42 &-47:27:04.45 &---- & ---  & $<$2.12 &63 &6042 &-1.72 & 0.02\\
240490 &13:26:23.83 &-47:26:42.34 &---- & ---  & $<$2.20 &50 &6042 &-1.51 & 0.26\\
240604 &13:26:25.92 &-47:26:40.71 &---- & ---  & $<$2.13 &53 &5911 &-1.68 & 0.19\\
241341 &13:26:46.69 &-47:26:29.98 &---- & ---  & $<$1.89 &79 &5858 &-1.91 & 0.37\\
241861 &13:26:22.08 &-47:26:23.51 &---- & ---  & $<$2.10 &59 &6111 &-1.86 & 0.00\\
242697 &13:26:56.94 &-47:26:11.44 &33.8 & 3.8  &    2.20 &59 &5961 &-1.42 & 0.23\\
242803 &13:26:23.68 &-47:26:10.98 &25.2 & 4.2  &    2.21 &54 &6171 &-1.63 & 0.28\\
244548 &13:26:41.42 &-47:25:47.46 &22.3 & 3.6  &    1.94 &62 &5878 &-1.72 & 0.24\\
245711 &13:26:22.65 &-47:25:31.92 &---- & ---  & $<$2.30 &50 &6107 &-1.56 & 0.23\\
246815 &13:27:08.72 &-47:25:15.22 &---- & ---  & $<$2.08 &47 &6008 &-1.59 & 0.16\\
246822 &13:26:43.28 &-47:25:16.07 &---- & ---  & $<$1.84 &51 &6011 &-1.71 & 0.21\\
246899 &13:26:57.38 &-47:25:14.46 &---- & ---  & $<$2.26 &59 &6127 &-1.46 & 0.28\\
246949 &13:26:59.34 &-47:25:13.73 &33.3 & 4.4  &    2.35 &51 &6176 &-1.72 & 0.21\\
247449 &13:26:36.70 &-47:25:07.74 &---- & ---  & $<$2.13 &61 &6012 &-1.68 & 0.17\\
248411 &13:26:51.38 &-47:24:54.65 &---- & ---  & $<$1.82 &58 &5939 &-1.61 & 0.14\\
248991 &13:26:54.85 &-47:24:46.72 &---- & ---  & $<$2.13 &40 &6032 &-1.54 & 0.24\\
249154 &13:26:56.74 &-47:24:44.10 &---- & ---  & $<$2.16 &58 &6023 &-1.68 & 0.28\\
249555 &13:26:24.96 &-47:24:39.03 &31.6 & 4.9  &    2.14 &46 &5926 &-1.57 & 0.13\\
249952 &13:26:28.89 &-47:24:33.44 &---- & ---  & $<$2.26 &48 &6121 & ---- & ----\\
250239 &13:26:49.98 &-47:24:28.90 &45.2 & 3.5  &    2.26 &65 &5838 &-1.71 & 0.17\\
250295 &13:26:47.08 &-47:24:28.29 &23.9 & 3.3  &    1.93 &68 &5823 &-1.75 & 0.14\\
250960 &13:26:41.05 &-47:24:19.16 &---- & ---  & $<$2.16 &67 &5934 &-1.55 & 0.21\\
300010 &13:25:42.83 &-47:36:09.89 &47.9 & 3.5  &    2.35 &64 &5925 &-1.78 & 0.22\\
300098 &13:25:52.22 &-47:36:06.94 &48.1 & 4.1  &    2.30 &55 &5852 &-1.75 & 0.22\\
300147 &13:25:47.93 &-47:36:05.63 &33.2 & 2.7  &    2.22 &83 &6006 &-1.87 & 0.37\\
311493 &13:25:45.82 &-47:31:39.66 &29.9 & 4.1  &    2.24 &54 &6107 &-1.89 & 0.38\\
608199 &13:26:39.18 &-47:41:25.72 &---- & ---  & $<$2.12 &45 &5991 &-1.69 & 0.15\\
608330 &13:26:41.25 &-47:41:21.00 &40.7 & 4.4  &    2.20 &51 &5832 &-1.45 & 0.16\\
608523 &13:26:38.02 &-47:41:13.32 &36.7 & 3.3  &    2.29 &68 &6021 &-1.94 & 0.20\\
\hline		
\end{tabular}	
\end{center}	
\end{table*}	
\addtocounter{table}{-1}
\begin{table*}
\caption{Basic data for MS/SGB stars studied in this paper (continued). For each star, we report ID, coordinates, the Li line EWs and errors, and the lithium abundances. The spectra signal-to-noise, H$_\alpha$-based effective temperatures, and the estimated [Fe/H] and its errors are also reported. For three stars, no [Fe/H] was estimated. These stars were not plotted in the bottom panel of Fig.\ref{figteff}.}
\begin{center}								     
\begin{tabular}{c|cc|ccr|c|c|cc}  					     
\hline									     
ID & RA  & Dec & EW & $\epsilon$ (EW) & A(Li) & S/N & \teff (K) & [Fe/H] & $\epsilon$ ([Fe/H])  \\ 	
\hline								             
610001 &13:26:39.48 &-47:40:20.13 &32.9 & 3.5  &    2.28 &65 &6088 &-1.69 & 0.26\\
610873 &13:26:46.88 &-47:39:51.34 &36.4 & 3.8  &    2.24 &59 &5972 &-1.54 & 0.15\\
611625 &13:26:33.36 &-47:39:29.26 &54.7 & 3.3  &    2.38 &68 &5870 &-1.62 & 0.17\\
612223 &13:26:52.72 &-47:39:11.66 &33.5 & 4.5  &    2.16 &49 &5917 &-1.51 & 0.31\\
612310 &13:26:39.46 &-47:39:09.45 &50.5 & 3.2  &    2.35 &71 &5886 &-1.88 & 0.10\\
612354 &13:26:24.63 &-47:39:08.21 &34.7 & 4.2  &    2.06 &53 &5763 &-1.71 & 0.28\\
613844 &13:26:32.11 &-47:38:29.44 &55.1 & 3.5  &    2.35 &65 &5816 &-1.87 & 0.24\\
614247 &13:26:23.67 &-47:38:19.62 &35.3 & 3.6  &    2.18 &62 &5900 &-1.86 & 0.30\\
614387 &13:26:27.86 &-47:38:16.08 &19.8 & 3.7  &    1.96 &61 &5976 &-1.78 & 0.16\\
614413 &13:26:50.27 &-47:38:15.14 &32.8 & 3.2  &    2.24 &69 &6031 &-1.76 & 0.21\\
614706 &13:26:22.07 &-47:38:08.14 &36.7 & 4.1  &    2.16 &54 &5854 &-1.66 & 0.20\\
615000 &13:26:35.56 &-47:38:01.05 &32.9 & 4.1  &    2.26 &55 &6058 &-1.72 & 0.37\\
615659 &13:26:25.62 &-47:37:46.49 &---- & ---  & $<$1.67 &56 &5921 &-1.58 & 0.23\\
615841 &13:26:31.83 &-47:37:42.53 &---- & ---  & $<$2.16 &63 &6103 &-1.56 & 0.20\\
617253 &13:26:53.48 &-47:37:12.63 &52.2 & 3.7  &    2.25 &61 &5725 &-1.95 & 0.27\\
617587 &13:26:40.17 &-47:37:06.10 &37.7 & 3.3  &    2.32 &68 &6057 &-1.87 & 0.21\\
619767 &13:26:38.98 &-47:36:24.68 &43.6 & 3.5  &    2.25 &63 &5863 &-1.87 & 0.13\\
706196 &13:26:18.98 &-47:41:06.91 &50.9 & 3.2  &    2.23 &70 &5728 &-1.82 & 0.24\\
707431 &13:26:10.29 &-47:39:57.26 &36.5 & 3.6  &    2.06 &62 &5722 &-1.69 & 0.13\\
707740 &13:26:17.15 &-47:39:41.58 &41.3 & 4.3  &    2.35 &52 &6023 &-1.48 & 0.12\\
709357 &13:26:13.14 &-47:38:31.22 &33.6 & 3.6  &    2.05 &63 &5771 & ---- & ----\\
709669 &13:26:01.52 &-47:38:18.12 &32.2 & 3.7  &    2.09 &61 &5843 &-1.96 & 0.15\\
710011 &13:26:11.96 &-47:38:05.21 &55.1 & 3.3  &    2.49 &68 &6003 &-1.75 & 0.38\\
710543 &13:25:58.10 &-47:37:44.77 &64.1 & 3.5  &    2.38 &64 &5749 &-1.80 & 0.10\\
711304 &13:26:19.02 &-47:37:16.13 &25.1 & 3.1  &    2.06 &72 &5960 &-1.94 & 0.19\\
711572 &13:25:55.45 &-47:37:07.35 &43.4 & 3.2  &    2.15 &71 &5732 &-1.86 & 0.18\\
711606 &13:26:03.26 &-47:37:06.37 &34.1 & 2.8  &    2.14 &80 &5878 &-1.96 & 0.16\\
\hline		
\end{tabular}	
\end{center}	
\end{table*}	

\begin{table*}
\caption{Basic data for SGB-a stars studied in this paper. For each star, we report ID, coordinates, and the derived upper limits to the lithium abundances. The spectra signal-to-noise and the H$_\alpha$-based effective temperatures are also reported. A metallicity of [Fe/H]=-1.5 was assumed.}\label{abuntablesgba}
\begin{center}								     
\begin{tabular}{c|cc|c|c|c}  					     
\hline									     
ID & RA  & Dec & A(Li) & S/N & \teff (K) \\ 	
\hline								             
120117 &13:27:15.18 &-47:29:52.13 & $<$1.50 &38 &5654\\
213295 &13:26:42.67 &-47:32:54.29 & $<$1.42 &32 &5490\\
213952 &13:26:23.23 &-47:32:45.23 & $<$1.23 &45 &5479\\
215700 &13:26:30.16 &-47:32:21.81 & $<$1.33 &44 &5542\\
216031 &13:26:27.23 &-47:32:17.22 & $<$1.00 &33 &5081\\
223861 &13:26:22.04 &-47:30:31.76 & $<$1.36 &51 &5330\\
224921 &13:27:08.47 &-47:30:15.98 & $<$1.39 &37 &5528\\
243327 &13:26:32.84 &-47:26:03.91 & $<$0.92 &46 &5148\\
247798 &13:26:31.46 &-47:25:03.30 & $<$1.30 &44 &5503\\
248814 &13:27:05.20 &-47:24:48.71 & $<$1.28 &43 &5465\\
\hline		
\end{tabular}	
\end{center}	
\end{table*}	

\end{document}